\documentclass{pazh}
\usepackage{graphicx}

\newcommand{\km}{\,\mbox{km}\,\mbox{s}^{-1}}
\def\Ha{\hbox{H$_\alpha$\,}}
\def\Hb{\hbox{H$_\beta$\,}}
\def\sig{\hbox{$\sigma_*$}}
\def\farcm{\hbox{$.\mkern-4mu^\prime$}}

\begin{document}

\title{Two-dimensional spectroscopy of double-barred galaxies}

\author{A.V. Moiseev}

\institute{  Special Astrophysical Observatory RAS, Nizhni\u{\i} Arkhyz,
Karachai--Cherkessia, 357147 Russia}

\offprints{Alexei Moiseev, \email{moisav@sao.ru}}
\date{Received May 29, 2002 }


\abstract{ We describe the results of our spectroscopy for a sample of barred
galaxies whose inner regions exhibit an isophotal twist commonly called a
secondary bar.  The line-of-sight velocity  fields of the ionized gas and stars
and the light-of-sight velocity dispersion  fields of the stars were
constructed from two-dimensional spectroscopy with the 6m Special Astrophysical
Observatory telescope. We detected various types of non-circular motions of
ionized gas: radial flows within large-scale bars, counter-rotation of the gas
and stars at the center of NGC~3945, a polar gaseous disk in NGC~5850, etc.
Our analysis of the optical and near-infrared images (both ground-based and
those from the Hubble Space Telescope) revealed circumnuclear minispirals in
five objects.  The presence of an inner (secondary) bar in the galaxy images
is shown to have no effect on the circumnuclear kinematics of the gas and
stars. Thus, contrary to popular belief, the secondary bar is not a dynamically
decoupled galactic structure. We conclude that the so-called double-barred
galaxies are not a separate type of galaxies but are a combination of objects
with distinctly different morphology of their circumnuclear regions. (c) 2002
MAIK Nauka/Interperiodica. }

\maketitle

\section{Introduction}

  According to observational
estimates, galaxies with central bars account for a major fraction (50-70\%)
of the total number of nearby disk galaxies (Selwood and Wilkinson, 1993;
Knapen et al.,  2000b).  The motion of stars and gaseous clouds within the bar
differs markedly from unperturbed circular rotation; the radial flows of gas
toward the center prove to be significant, as confirmed by direct observations
(Afanasiev and Shapovalova, 1981; Duval and Monnet, 1985; Knappen et al.,
2000a) and by numerous model calculations [see Lindblad (1999) for a review].
The central regions of such galaxies are decoupled in their dynamical
parameters, star-formation rates, and densities of the gas and dust.  For
example, the molecular-gas density within the central kiloparsec in barred
galaxies is an order of magnitude higher than that in unbarred galaxies
(Sakamoto et al. 1999).

The dynamical effect of the bar is considered to be a major mechanism of
transporting interstellar gas from the disk into the circumnuclear region,
where it becomes fuel for a circumnuclear starburst or an active nucleus
(Combes, 2001).  In the latter case, however, the relationship between the bar
and the active (for a disk galaxy, Seyfert) nucleus is far from being
unequivocal.  Thus, the relative fraction of bars  in Seyfert galaxies only
slightly exceeds this fraction in galaxies without active nuclei (Knapen et
al., 2000).  The main problem is that the gas in the bar is concentrated not
in the nucleus itself but in a ring of several hundred parsecs in radius, in
the region of the inner Lindblad resonance.  Therefore, an additional
mechanism is required to take angular momentum away from the gas at a distance
of $100-1000$\,pc from the centre and to transport the gas into the region
where the gravitational forces of a central supermassive black hole are in
action (Combes, 2001).  An elegant solution to the problem of mass transport to
an active nucleus is the assumption made by Shlosman et al. (1989) that another
bar can be formed in the gaseous disk (ring) within a large-scale bar that
again produces flows of gas toward the nucleus.  The system of two bars is
capable of sweeping away the interstellar medium on scales of several kpc and
of concentrating it at distances of $1-10$ pc from the centre.  Recently, this
process was numerically simulated by Heller et al. (2001), although, in
essence, the authors considered the evolution of the inner elliptical ring
rather than the stellar-gaseous bar.

Increased interest in double bars stems from the fact that something like that
is occasionally seen in the images of barred galaxies.  Vaucouleurs (1975)
detected a bar-like structure within the large-scale bar in NGC~1291.  The
first systematic observational study of a double-barred galaxy was undertaken
by Buta and Crocker (1993).  They published a list of 13 galaxies with an
arbitrary orientation of the inner (secondary) bar relative to the outer
(primary) bar. Analysis of the isophotes shapes revealed secondary bars in the
optical (Wozniak et al., 1995; Erwin and Sparke 1999) and NIR images of
galaxies (Friedli et al., 1996; Jungwiert et al., 1997; Laine et al., 2002). An
extensive bibliography and a list of 71 candidates for double bars compiled
from literature are given in Moiseev (2001b).

\begin{figure}
\includegraphics[width=8  cm]{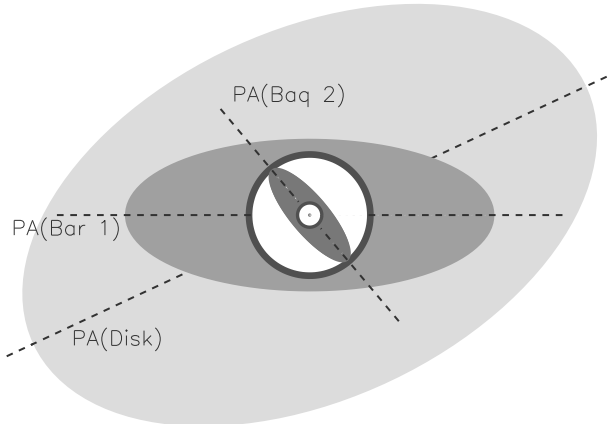}
 \protect\caption{
A schematic view of a double-barred galaxy in projection onto the plane of the
sky.  The ring that corresponds to the region of the inner Lindblad resonance
in the outer bar is highlighted.  The principal isophotal orientations (see
Subsect.~3.1) are indicated by dashed lines. }
\end{figure}

Although there are now many papers on this subject, the dynamical behavior of
such stellar configurations is still unclear.  Maciejewski and Sparke (2000)
showed that closed orbital loops maintaining the shapes of both bars that
rotate with different angular velocities could exist.  Similar independently
rotating structures also occasionally appear in experiments on simulating
stellar-gaseous disks (Pfenninger and Norman, 1990; Friedli and Martinet,
1993).  The behavior of the gas in double bars was numerically analyzed by
Maciejewski et al., (2002) and Shlosman and Heller (2002).  Figure 1 shows a
scheme of a galaxy with two independently rotating bars, which may actually be
considered to be universally accepted.

It should be noted that, despite several interesting results obtained in
numerical experiments, they are strongly model-dependent. Khoperskov et al.,
(2001) showed that a secondary bar could periodically arise only at certain
stages of the galaxy dynamical evolution; a long-lived secondary bar cannot
yet be simulated (Friedli and Martinet, 1993; Erwin and Sparke, 2002).  New
observational data are required to verify contradictory theoretical
predictions.

Numerous observational studies indicate that in the case of double bars, we
probably come across a new structural feature of barred galaxies.  However,
the vast majority of these studies are based only on photometric data, when an
extended structure is seen in the image of the galaxy within its primary bar
(Fig. 2).  Formal application of an isophotal analysis (Wozniak et al., 1995)
even allowed several authors to distinguish triple bars (Jungwiert et al.,
1997; Friedli et al., 1996; Erwin and Sparke, 1999) without any reasoning on
the dynamical stability of such configurations. However, the observed
photometric structural features of such galaxies can also be explained in less
exotic ways, without invoking secondary or third bars.  An oblate bulge, an
intricate distribution of star-forming regions and dust in the circumnuclear
region, and an elliptical ring in the resonance region of the major bar can
all create the illusion of a secondary bar in the galaxy images (Friedli et
al., 1996; Moiseev, 2001b).  Kinematic data, i.e. , measurements of the
line-of-sight velocities and velocity dispersions of the gas and stars, are
required to solve the problem.  Since the observed objects are definitely not
axisymmetric, two-dimensional spectroscopy can be of great help.  It allows
the two-dimensional distributions of line-of-sight velocities and their
dispersions in the plane of the sky to be constructed.

In this paper, we discuss the results of the first systematic study of such
galaxies by using two-dimensional spectroscopy carried out in 2000-2002 with
the goal to answer the following question: Are the secondary bars dynamically
decoupled systems? Since the observational data themselves were described in
detail by Moiseev et al. (2002), we discuss here only the most important
features of the objects under study.  The observing techniques are described
in Sect.~2. In Sect.~3, we analyze the velocity and velocity dispersion
distributions in the galaxies and describe the minispiral structures detected
in the central regions of several galaxies.  Our results are discussed in
Sect.~4 and our main conclusion is formulated in Sect.~5.

\section{Observations and data reduction}

We drew our sample from the list of candidates for double bars (Moiseev, 2001b)
based on convenience of their observation with the 6m telescope: $\delta>0$;
the diameter of the secondary bar fits into the MPFS field of view.
Observational data were obtained for 13 galaxies, which account for about half
of the total number of such objects in the northern sky.  The Table~1 gives the
name of the galaxy, its morphological type from the NED database, and the
sizes of the apparent semi-major axes of the outer ($a_1$) and inner ($a_2$)
bars in arcseconds with a reference to the corresponding papers.

All spectroscopic and some of the photometric observations were carried out
with the 6m Special Astrophysical Observatory (SAO, Russia) telescope at $1-2.5''$ seeing.
The detector was a TK1024 CCD array.  A log of observations and a detailed description of
individual galaxies and the data reduction procedure were given by Moiseev et al. (2002).

\begin{table*}[t]
\caption{ Parameters of the observed galaxies}
\begin{tabular}{|l|l|r|r|l|}
\hline
Name &Type & $a_1\,('')$ & $a_2\,('')$ & References to photometry \\
\hline
NGC  470 &SAb   &32 &8  &Wozniak et al. (1995); Friedli et al. (1996)\\
NGC 2273 &SBa   &24 &8  &Mulchaey et al. (1997)\\
NGC 2681 &SAB0/a&29 &5  &Wozniak et al. (1995); Erwin and Sparke (1999)\\
NGC 2950 &SB0   &38 &6  &Wozniak et al. (1995); Friedli et al. (1996)\\
NGC 3368 &SABab &24 &4  &Jungwiert et al. (1997)\\
NGC 3786 &SABa  &25 &7  &Afanasiev et al. (1998)\\
NGC 3945 &SB0   &42 &20 &Wozniak et al. (1995); Erwin and Sparke (1999)\\
NGC 4736 &SAab  &26 &10 &Shaw et al. (1993)\\
NGC 5566 &SBab  &24 &6  &Jungwiert et al. (1997)\\
NGC 5850 &SBb   &84 &9  &Buta and Crocker (1993);Wozniak et al. (1995)\\
NGC 5905 &SBb   &37 &6  &Wozniak et al. (1995); Friedli et al. (1996)\\
NGC 6951 &SABbc &44 &5  &Wozniak et al. (1995)  \\
NGC 7743 &SB0   &57 &10 &Wozniak et al. (1995)   \\
\hline
\end{tabular}
\end{table*}

The circumnuclear regions of all galaxies were observed with the
Multipupil Field Spectrograph (MPFS) (Afanasiev et al., 2001).  It
simultaneously takes spectra from 240 spatial elements in the shape of square
lenses that comprise a $16\times15$ matrix in the plane of the sky.  The angular
size of a single matrix element is $1''$. The MPFS spectrograph is described in
the Internet at \verb*"http://www.sao.ru/~gafan/devices/mpfs/mpfs_main.htm"
The observations were carried out in the spectral range $\lambda 4800-6100$\AA\, and,
for several galaxies, in the range $\lambda5800-7100$\AA\,; the dispersion was
1.35\AA\, per pixel.  The covered spectral range included absorption features
typical of the old (G-K-type) galactic stellar population.  The line-of-sight
velocity and velocity dispersion fields of the stars were constructed by using
a cross-correlation technique modified for work with MPFS data (Moiseev, 2001a).
The line-of-sight velocities and velocity dispersions were determined with an
accuracy of $\sim10\km$.  Based on the MPFS observations, we
also mapped the two-dimensional intensity distribution and the velocity field of
the ionized gas in the \Hb, [O~III]$\lambda 4959, 5007$\AA\, and [N~II]$\lambda6548, 6583$
\AA\, emission lines.  The line-of-sight velocities were measured with an accuracy
of $\sim10 \km$.  No emission features were detected in NGC~2950 and NGC~5566.
Six galaxies with intense emission features were observed with a scanning
interferometer Fabry-Perot  (IFP) in the 235th order of interference in a
spectral region near the wavelength of the \Ha line.  The instrumental profile
width was $2.5$\AA\, ($\sim110 \km$); the field of view was about $5'$ with a scale
of $0.56-0.68''$ per pixel.  The instrument and reduction techniques were
described previously (Moiseev, 2002).  We constructed the velocity fields of the
ionized gas in \Ha or [N II]$\lambda 6583$\AA\, with an accuracy of $\sim5 \km$.

The optical V- and R-band images of seven galaxies were obtained at the prime
focus of the 6m telescope using the SCORPIO focal reducer (its description can
be found in the Internet at
\verb*"http://www.sao.ru/~moisav/scorpio/scorpio.html"). The field of view is
4\farcm8 with a scale of $0.28''$ per pixel. In addition, we used the JHK-band
images obtained with the 2.1m OAN telescope in Mexico [for more detail, see
Moiseev et al., (2002)].  We also used the galaxy images from the Hubble Space
Telescope (HST) archive obtained with the Wide-Field and Planetary Camera
(WFPC2) and with the Near-IR Camera (NICMOS). As an example of the
observational data used, Fig.~2 shows the images of NGC~2273 and NGC~2950 and
Fig.~3 shows the velocity fields of the gas and

\section{Analysis of the spectroscopic and photometric observations}

\subsection{Velocity dispersion}

The velocity dispersion distribution of the stellar disk is one of  its
important parameters.  It allows unambiguous multicomponent models of the mass
distribution in galaxies to be constructed (Khoperskov et al., 2002).
Numerical calculations show that because of the bar formation, the velocity
dispersion distribution in the galactic disk differs greatly from the
unperturbed (without a bar a nd spiral structure) axisymmetric case (Miller and
Smith, 1979; Vauterin and Dejonghe, 1997).  The bar is a much hotter dynamical
subsystem; the velocity dispersion in it increases sharply.  In addition, the
velocity ellipsoid is found to be highly anisotropic.  This anisotropy
manifests itself in different distributions of the radial, azimuthal, and
vertical velocity dispersions in the disk plane.  The model maps of the
line-of-sight stellar velocity dispersion constructed for various disk and bar
orientations (Miller and Smith, 1979; Vauterin and Dejonghe, 1997; Khoperskov
et al., 2001) indicate that the line-of-sight velocity dispersion (\sig)
distribution within the bar is symmetric about the bar major axis rather than
about the disk major axis (line of nodes), as would be the case in the absence
of a bar.  Unfortunately, the observational manifestations of the
velocity-dispersion anisotropy in the \sig distribution are few in number.  The
series of papers by Kormendy (1982, 1983) are an example of the most
consistent approach to measuring the velocity-dispersion anisotropy in barred
galaxies.  However, most authors restrict themselves to measuring \sig along
one or two spectrograph slit directions.  The two-dimensional \sig maps used
here are much more informative. Figures 3c and 3d show the isolines of the \sig
distribution that form ellipsoidal, elongated structures asymmetric about the
disk line of nodes in the central regions of NGC~2273 and NGC~2950.

\begin{figure*}
\includegraphics[width=16  cm]{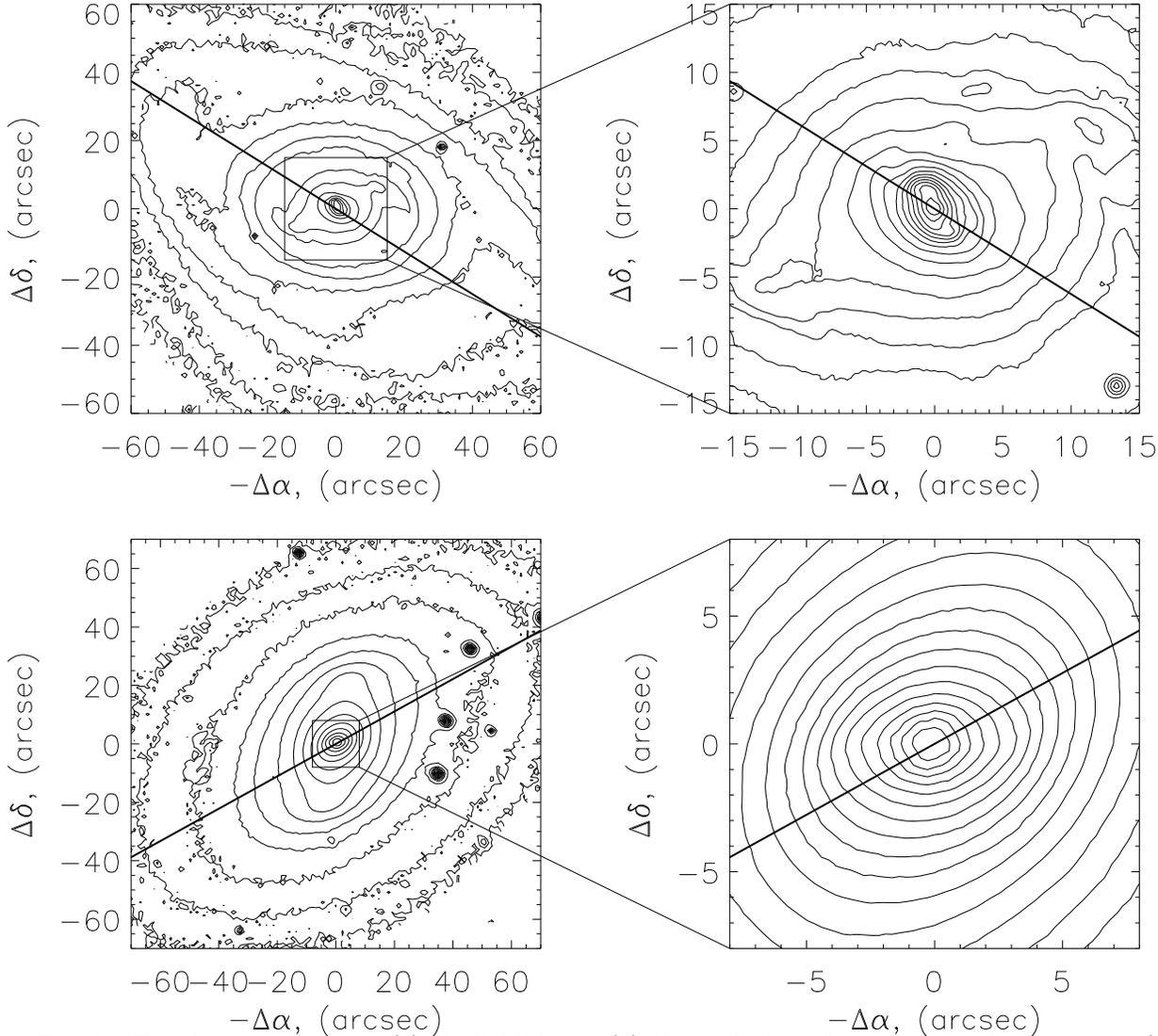}
 \protect\caption{
 The R-band images of (a) NGC~2273 and (b)
NGC~2950 obtained with the 6m telescope; (c) and (d) the enlarged central
regions highlighted by the square in panels (a) and (b), respectively. The
inner isophotal twist relative to the outer bars is clearly seen.  The heavy
lines indicate the orientation of the line of nodes of the disk.  stars and
the velocity dispersion fields of the stars in NGC~2273, NGC~2950, and NGC
3945.
}
\end{figure*}

 To quantitatively describe the deviation of the \sig
distribution from the axisymmetric case, we used the Fourier expansion of the
observed velocity dispersion field in terms of position angle $PA$:

\begin{equation}
\sigma_*(r,PA)=A_0(r)+\sum_{m=1}^{N}A_m(r)\cos(m\,PA+\phi_m(r)),
\end{equation}

 where r is the distance from the
center in the plane of the sky; $A_m$ and $\phi_m$ are the amplitude and phase of the
harmonic with number $m$, respectively; and $N = 6-8$ is the maximum number of
harmonics.  Our technique is similar to that used by Lyakhovich et al. (1997)
to analyze the velocity fields but differs from it in that the Fourier expansion
is made in terms of PA rather than in terms of the azimuthal angle in the
galactic plane.  In addition, it was shown in the above paper and in
subsequent papers (see, e.g., Fridman et al., 2001) that for the velocity
field of the gaseous disk in a spiral galaxy, the principal expansion harmonics
are related to the spatial velocity vector components.  For analysis of the \sig
fields, we cannot yet offer such a clear physical interpretation of the Fourier
spectrum,  because the combined contribution of the bulge and the disk
with the bar is observed in the central regions under study.  However, since
our simulations indicate that the \sig isolines in the bar form an ellipsoidal
structure, this must correspond to a situation where the $m = 2$ harmonic and,
possibly, the succeeding even harmonics have a maximum amplitude in (1).  The
direction of the major axis of this structure corresponds to the line of
maximum of the second harmonic: $PA_2=-\phi_2/2\pm180^\circ$.

\begin{figure*}
\includegraphics[width=16  cm]{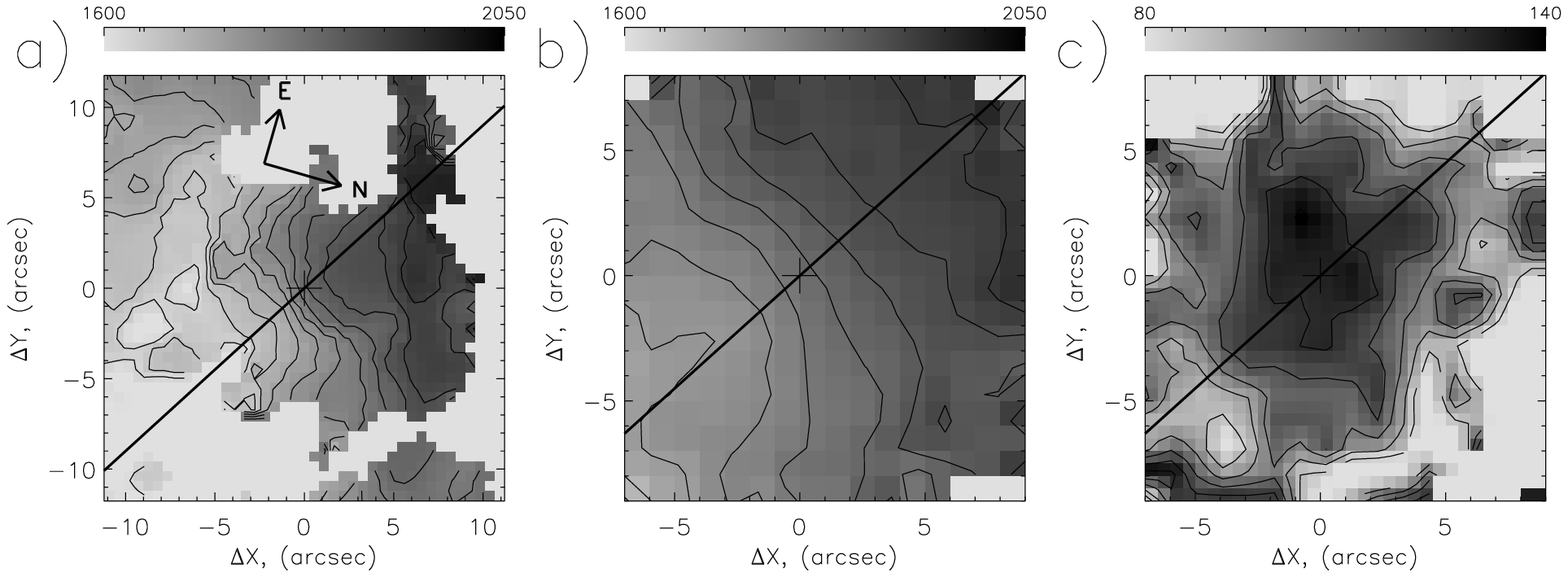}
\includegraphics[width=16  cm]{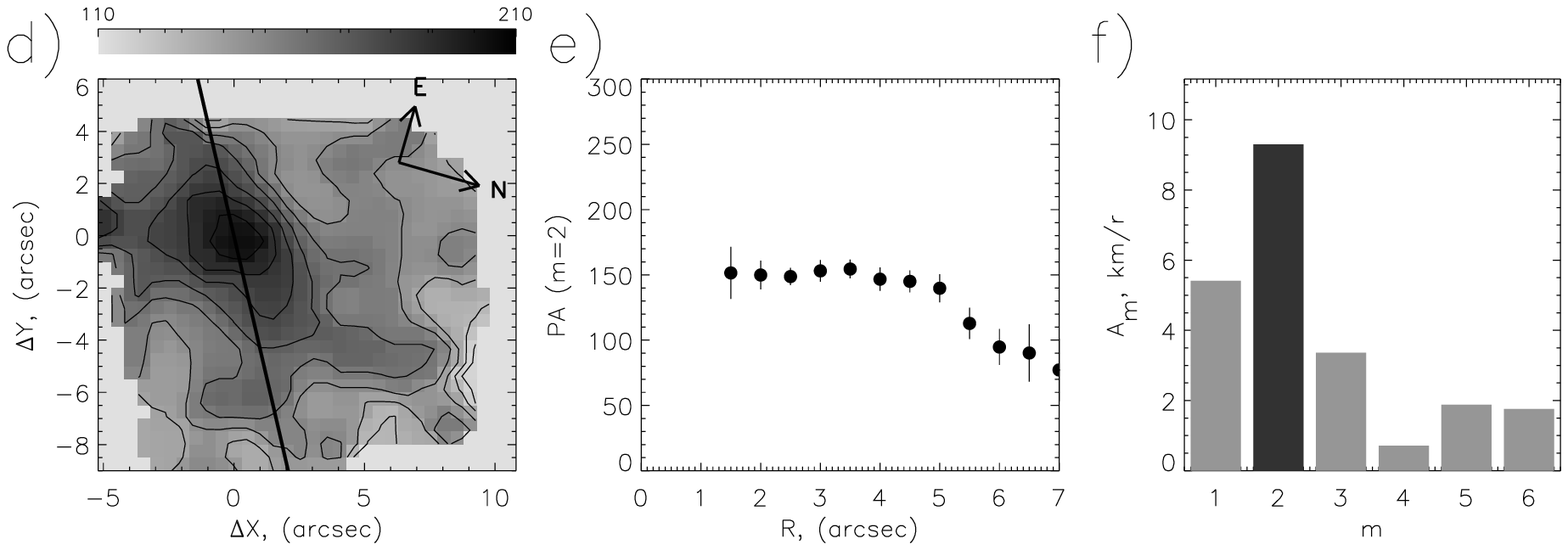}
\includegraphics[width=16  cm]{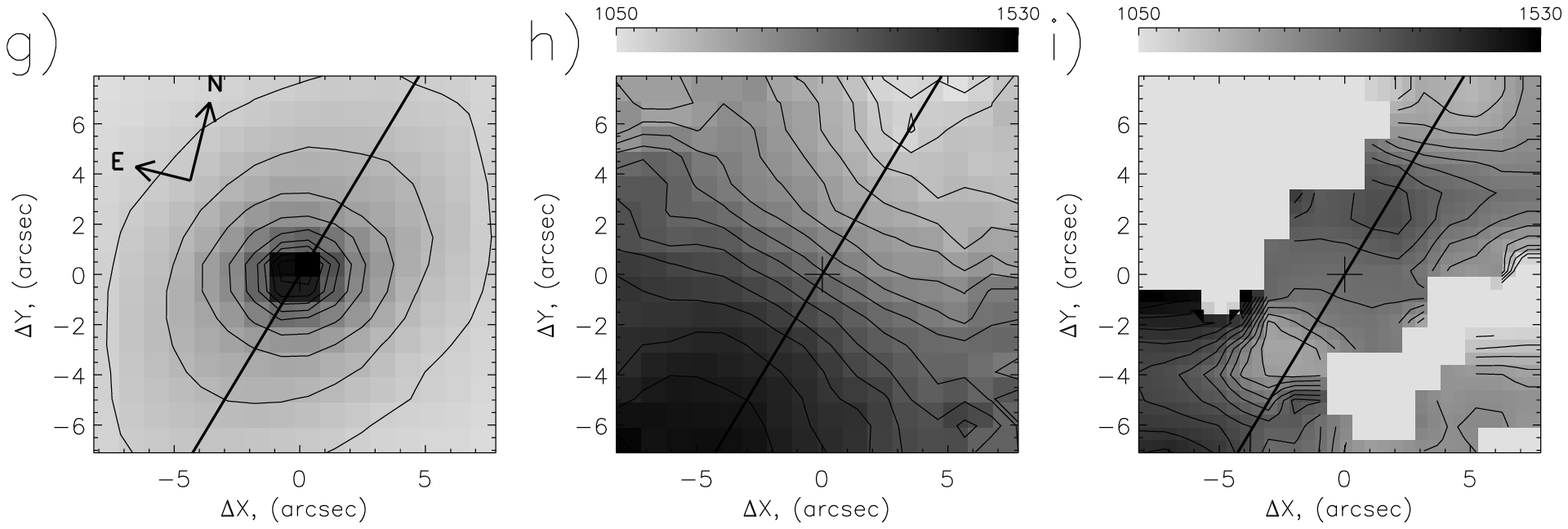}
 \protect\caption{
Results of the observations
of three galaxies.  The gray scale is velocity in $\km$ and the heavy line
indicates the orientation of the line of nodes everywhere.  NGC~2273: (a) the
gas velocity field constructed from IFP observations; (b), (c) the
stellar velocity and velocity dispersion fields.  NGC~2950: (d) the velocity
dispersion field; (e) the radial dependence of the position angle of the
maximum of the second harmonic in the Fourier spectrum $\sigma_*$; (f) the mean
harmonic amplitude in the Fourier spectrum $\sigma_*$.  NGC~3945: (g) the continuum
MPFS image of the center; (h) the stellar velocity field; (i) the velocity
field of the gas with counter-rotation in the central region.
}
\end{figure*}

We broke down the velocity dispersion fields into rings with the center
coincident with the photometric center of the continuum image.  To provide a
sufficient number of points in relation (1), each image element was broken
down into four $0.5\times 0.5''$ elements.  Experiments with analysis of
various images show that this procedure introduces no significant distortions
into the spectrum of the first harmonics, at least for $m < 4-5$.  The
position angle of the symmetry axis in the velocity dispersion distribution is
defined as the mean value of the $r$ dependence of $PA_2$, provided that there
is a segment with an approximately constant $PA_2$ and that the second
harmonic dominates in the Fourier spectrum at given r.  For NGC~2950, this
radii range is $r = 1-5''$ (Figs.  3e, 3f).  The succeeding change in $PA_2$
at large distances stems from the fact that the galaxy does not lie exactly at
the center of the MPFS field of view (Fig.  3d). Therefore, there are few
pairs of diametrically opposite points in the velocity dispersion field for $r
> 5''$, causing the spectrum of the even harmonics to be distorted.

\begin{figure*}
\includegraphics[width=16  cm]{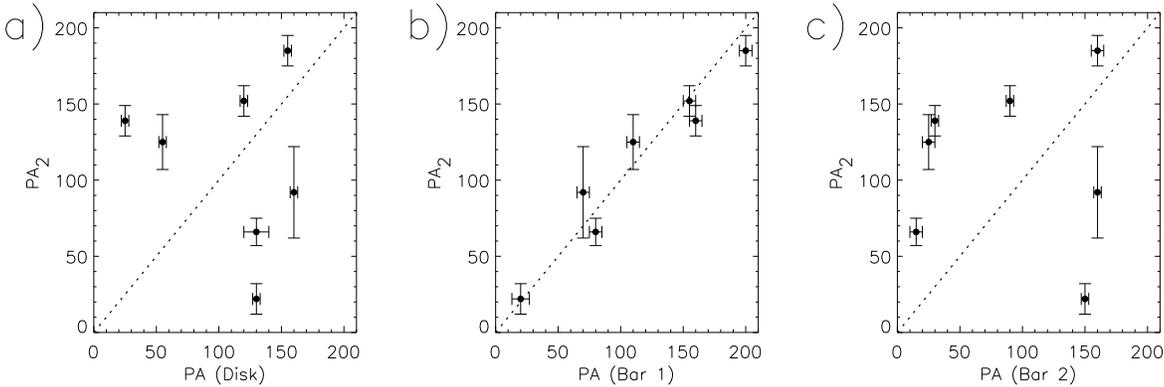}
 \protect\caption{Relation between the position
angle of the second harmonic in the velocity dispersion distribution and the
position angle of the disk (a) and the two bars (b) and (c). The dotted line
represents a direct proportionality of PAs (not the best fit!) }
\end{figure*}

The following three principal isophotal
orientations can be distinguished in the image of a double-barred galaxy: the
position angle of the disk, $PA$(Disk), and the position angles of the bars,
$PA$(Bar~1) and $PA$(Bar~2), shown in Fig.~1. According to the above
considerations, if there are two dynamically independent bars, then each of
these directions must be the symmetry axis in the apparent \sig distribution on
the corresponding distance scale.  What actually determines the velocity
dispersion distribution in the central region? To answer this question, we
considered the relations between the position angle $PA_2$ of the symmetry axis
of the velocity dispersion field and these three directions shown in Fig.~4.
This figure shows data only for those seven objects from our sample in which the
second harmonic dominates in the Fourier spectrum of the azimuthal \sig
distribution at $r = 1-6''$.  We see that in the galaxies under study, the
direction $PA_2$ in the circumnuclear region ($r < 5-6''$) coincides (within the
error limits) only with the orientation of the outer bar and correlates
neither with the major axis of the inner bar (Fig.~4c) nor with the line of
nodes of the disk (Fig.~4a).

Since the central regions in galaxies of mostly early morphological types were
observed (see the Table~1), bulge stars must contribute significantly to the
velocity dispersion. However, if the bulge is spherical, then it will affect
only the amplitude of the $m=0$ harmonic, because the expansion (1) is made in
terms of the angle in the plane of the sky.  If the bulge is oblate but
axisymmetric (spheroidal), then this will cause an increase in the amplitude
of the second harmonic.  The line of its maximum must coincide with the line
of nodes of the disk, because the symmetry of the observed \sig distribution
is similar to the case of a disk with a different apparent axial ratio.
However, Fig.~4a shows no correlation between $PA_2$ and the line of nodes.
If, alternatively, the bulge is triaxial, then it will produce an isophotal
twist in the central region in projection onto the plane of the sky (Wozniak
et al., 1995) and will be barely distinguishable from the inner bar.  However,
Fig.~4c shows no correlation between the symmetry direction of the velocity
dispersion and the inner isophotal orientation.

Thus, the location of the line of maximum of the second
harmonic in the Fourier expansion of the velocity dispersion field correlates
only with the outer bar.  This large-scale bar determines the dynamics of the
stellar component even in those regions where the isophotal twist attributed
to the secondary bar is observed.

\subsection{Velocity  fields}

We determined the radial dependences of the position angle of the dynamical
major axis (the line of maximum line-of-sight velocity gradient) from the
velocity field by the method of inclined rings (Begeman, 1989).  The velocity
fields were broken down into elliptical rings about $1''$ in width aligned with
the outer galactic disk.  In each ring, we determined the optimum dynamical
position angle $PA_{dyn}$\, in the approximation of circular rotation [for more
details, see Begeman, 1989; Moiseev and Mustsevoi, 2000].

The gaseous clouds move within the bar in such a way that the observed
$PA_{dyn}$\, ceases to be aligned with the line of nodes of the disk
(Chevalier and Furenlid, 1978); the dynamical axis turns in a sense opposite to
the line of nodes compared with the position angle of the inner isophotes
(Moiseev and Mustsevoi, 2000).  In other words, the lines of equal
line-of-sight velocities seek to elongate along the bar.  A similar effect
must also be observed in the stellar velocity field, as shown both by numerical
simulations (Vauterin and Dejonghe, 1997) and by analytic calculations of the
stellar dynamics in a triaxial gravitational potential (Monnet et al., 1992).

Figure~5 shows the radial dependences of the position angles of the dynamical
axis and the major axis of the inner isophotes in several objects.  We
determined the isophotal orientation in NGC 2950 and NGC~3786 using the HST
images obtained with the WFPC2 camera through a F814W filter and with the
NICMOS~1 camera through a F160W filter, respectively. The R-band image of
NGC~5850 was taken from the digital atlas by Frei et al. (1996).  In NGC~2950,
the isophotal orientation changes by more than $50^\circ$  (Fig.~5a), but
$PA_{dyn}$\, for the stellar component is virtually aligned with the line
 of nodes.  Neither the outer bar nor the inner bar have an appreciable effect on
the stellar velocity field.  This is probably because here, both stellar
motions within the bar and the rotation of the bulge, whose contribution in
this lenticular galaxy must be significant, are observed along the
line-of-sight. Since the spectral resolution was too low to separate the two
dynamical components in the line-of-sight velocity distribution (LOSVD), the
mean velocity field corresponds to circular rotation. As was noted in
Subsect.~3.1, the stellar component associated with the outer bar in NGC~2950
shows up in the distribution of the velocity dispersion, whose increase points
to a broadening of the LOSVD in the bar region.

 A similar pattern of stellar
motion is observed in NGC~2273, NGC~3945, NGC~5566, NGC~5850, and NGC~5905. In
four galaxies (NGC~470, NGC~2681, NGC~4736, and NGC~6951), there are
significant ($7-20^\circ$) deviations of $PA_{dyn}$\, for stars from the line
of nodes of the outer disk; these deviations of the dynamical axis are not
related to the orientation of the inner bar but are exactly in the opposite
direction compared with the isophotes of the outer bar.  Here, as with the
velocity dispersion, the stellar motions are affected by the outer bar rather
than the inner bar, as should be the case for the model of independently
rotating bars described in Sect.~1.

In NGC~3786, on the scale of the inner bar ($r < 6''$), $PA_{dyn}$\, deviates
from the line of nodes by more than $10^\circ$; the deviations are in the
opposite direction compared with the central isophotes (Fig.~5b). Here, we
actually observe a dynamically decoupled central minibar about $\sim2$~kpc in
diameter. Its existence was first suspected by Afanasiev and Shapovalova (1981)
when studying the gas kinematics in NGC~3786.  In this galaxy, however, we
cannot speak about a dynamically decoupled secondary bar, because a Fourier
analysis of the surface-brightness distribution (Subsect.~3.3) indicates that
the two-armed spiral mistaken by Afanasiev et al., (1998) for the outer bar is
located in NGC~3786 at $r = 7-20''$.

The situation in NGC~3368 is similar. A dynamically decoupled inner bar is
observed in this galaxy, but the outer bar described by Jungwiert et al.,
(1997) is actually part of the spiral structure. The paper that describes in
detail the structure and dynamics of NGC~3368 is being prepared for
publication (Sil'chenko et al., 2003).

\begin{figure}
\includegraphics[width=8 cm]{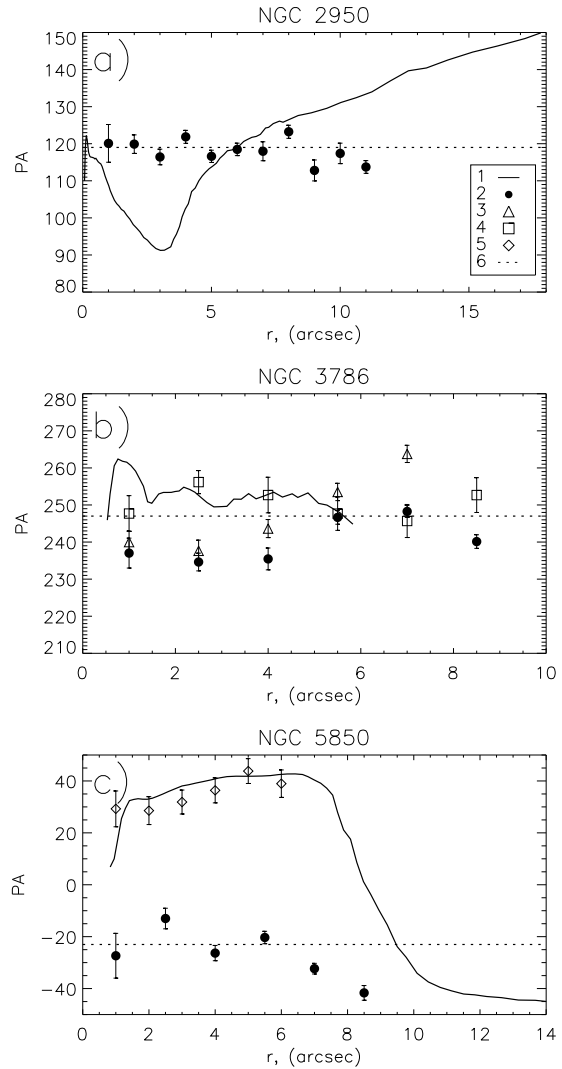}
 \protect\caption{
 The radial behavior of the position angle of the inner isophotes (1) and of
the dynamical axis determined from stars (2) and from gas in the H$_\beta$
(3), [O~III] (4), and [N~II] (5) lines for (a) NGC~2950, (b) NGC 3786, and (c)
NGC~5850; 6, the line of nodes of the outer disk. }
\end{figure}

  The apparent kinematics  of the ionized gas and stars in the objects under study are often
different. Moreover, the ionized-gas velocities measured from permitted (\Ha,
\Hb) and forbidden ([O~III], [N~II]) lines can differ markedly.  In NGC~3786,
the dynamical axes in the stellar  and gas velocity fields in \Hb are aligned,
while the position angles measured from the velocities in the [O~III] line
systematically deviate from them (Fig.  5b). A similar effect is also observed
in NGC~470, NGC~2273, NGC~5905, and NGC~6951. These deviations may be due to
the presence at the bar edges of shock fronts  formed in the disk gas under
action of the with the bar gravitational potential.  The post-shock gas
decelerates and emits in forbidden lines (Afanasiev and Shapovalova, 1981). The
following alternative explanation is also possible: neglected stellar
absorption features distort the \Hb emission profile, which, in turn,
introduces systematic errors into the gas line-of-sight velocity measurement.
However, in galaxies with relatively intense emission features against the
background of low-contrast absorption features, the velocity difference in
forbidden and permitted lines in the bar region can be real.

In all the sample galaxies in which intense emission features were observed,
the position angle $PA_{dyn}$\, measured from the gas is misaligned with the
line of nodes of the outer disk, suggesting non-circular motions within the
central kiloparsec.  In five galaxies, the turn of the dynamical axis is not
related to the isophotal twist in the inner bar. At the same time, it is in the
opposite direction relative to the isophotes of the outer bar.  This implies
that the outer bar determines the dynamics of the gas component, while the
inner bar is not dynamically decoupled.

The non-circular gas motions in the remaining six galaxies are different in
nature.  Thus, in NGC~3368 and NGC~3786, they are associated with the central
minibar and there is no outer bar here, as was mentioned above. In NGC~6951,
the non-circular gas motions are associated with the central minispiral, which
is discussed in the next section.

In NGC~470, the dynamical center, defined as the symmetry point of the velocity
field, is displaced by $4-5''$ ($0.6-0.8$~kpc) from the photometric nucleus of
the galaxy, which may be due to the peculiar development of the azimuthal $m=1$
harmonic in the gas velocity field (Emsellem, 2002).  This asymmetric harmonic
can be generated by the tidal interaction with the nearby companion NGC~474
studied by Turnbull et al. (1999).  The observed displacement of the center
can also be explained as resulting from the development of an asymmetric
harmonic in the surface brightness distribution. As was pointed out by Zasov
and Khoperskov (2002), this effect can be observed at certain evolutionary
stages of barred galaxies.

 The most peculiar galaxies (in terms of the gas
kinematics) are NGC~3945 and NGC~5850.  In the NGC~3945 the gas line-of-sight
velocities within $6''$ (0.5 kpc) are close in maximum amplitude to the stellar
velocities (80 and 120 $\km$, respectively) but are opposite in sign! At large
distances, the sense of rotation of the gas changes sharply and virtually
coincides with that of the stars (Fig.~3h,i).  The latter fact is also
confirmed by the line-of-sight velocity measurements of the ionized gas with
the IFP in several star-forming regions in the outer ring structure at
distances $120-140''$ (10-11 kpc) from the center.  It should be noted that
such counter-rotation of the gas and stars is occasionally observed in
early-type galaxies; it is commonly attributed to the absorption of an outer
gas cloud (Bertola et al., 1992; Kuijken et al., 1996).

The dynamical axis of the stars in the circumnuclear region of NGC~5850 is
close to the line of nodes, while in the ionized gas, this axis deviates from
it by more than $50-60^\circ$ (Fig.~5c) and is almost aligned with the position
angle of the central isophotes.  This behavior is typical of a disk inclined
to the galactic plane (Moiseev and Mustsevo, 2000).  In addition, if we,
nevertheless, assume the gas motions to take place in the galactic plane, then
it will turn out that they correspond to a radial outflow from the
nucleus\footnote{Here, we use the suggestion made by Higdon et al. (1998) that
the disk orientation of the galaxy for which its western half is closest to
the observer is most probable.} with velocities $50-70 \km$.  Such features are
characteristic of Seyfert galaxies, but the optical spectrum of the galaxy
contains no emission lines indicative of an active nucleus, nor are staburst
observed here (Higdon et al., 1998). A more reasonable assumption is that the
gas moves at $r < 6-7''$ in a plane polar to the galactic disk.  In this case,
the polar gaseous disk lies almost exactly along the small cross section of
the outer bar.  In recent years, such polar minidisks associated with a
large-scale bar or a triaxial bulge have been detected in the circumnuclear
regions of several galaxies, for example, in NGC~2841 (Sil'chenko et al. 1997;
Afanasiev and Sil'chenko, 1999) or NGC~4548 (Sil'chenko, 2002).  The
hypothesis of a polar disk is also supported by the fact that, according to
Higdon et al., (1998), NGC~5850 has undergone a recent collision with the
nearby galaxy NGC~5846. Through their interaction, part of the gas could be
transported to polar orbits.

\subsection{Minispirals}

 We used the HST archival
images of the galaxies to study the detailed morphology of their circumnuclear
regions.  The models of mean elliptical isophotes were constructed by the
standard technique and were subsequently subtracted from the original images.
In five galaxies, the residual brightness distributions obtained in this way
within their large-scale bars are in the shape of small spirals $5-15''$ in
size.  A Fourier analysis of the azimuthal brightness distribution was used to
quantitatively describe the detected spirals.  The original images at each
radius were expanded into the Fourier series (1), with the radius and the
azimuthal angle in the galactic disk plane being applied for $r$ and $PA$,
respectively.  The derived Fourier spectrum allows us to determine both the
number of arms of the principal spiral harmonic and the location of the line
of its maximum amplitude.  Examples   of spirals, their residual brightness,
and the mean harmonic amplitudes in three galaxies are shown in Fig.~6.

\begin{figure*}
\includegraphics[width=15 cm]{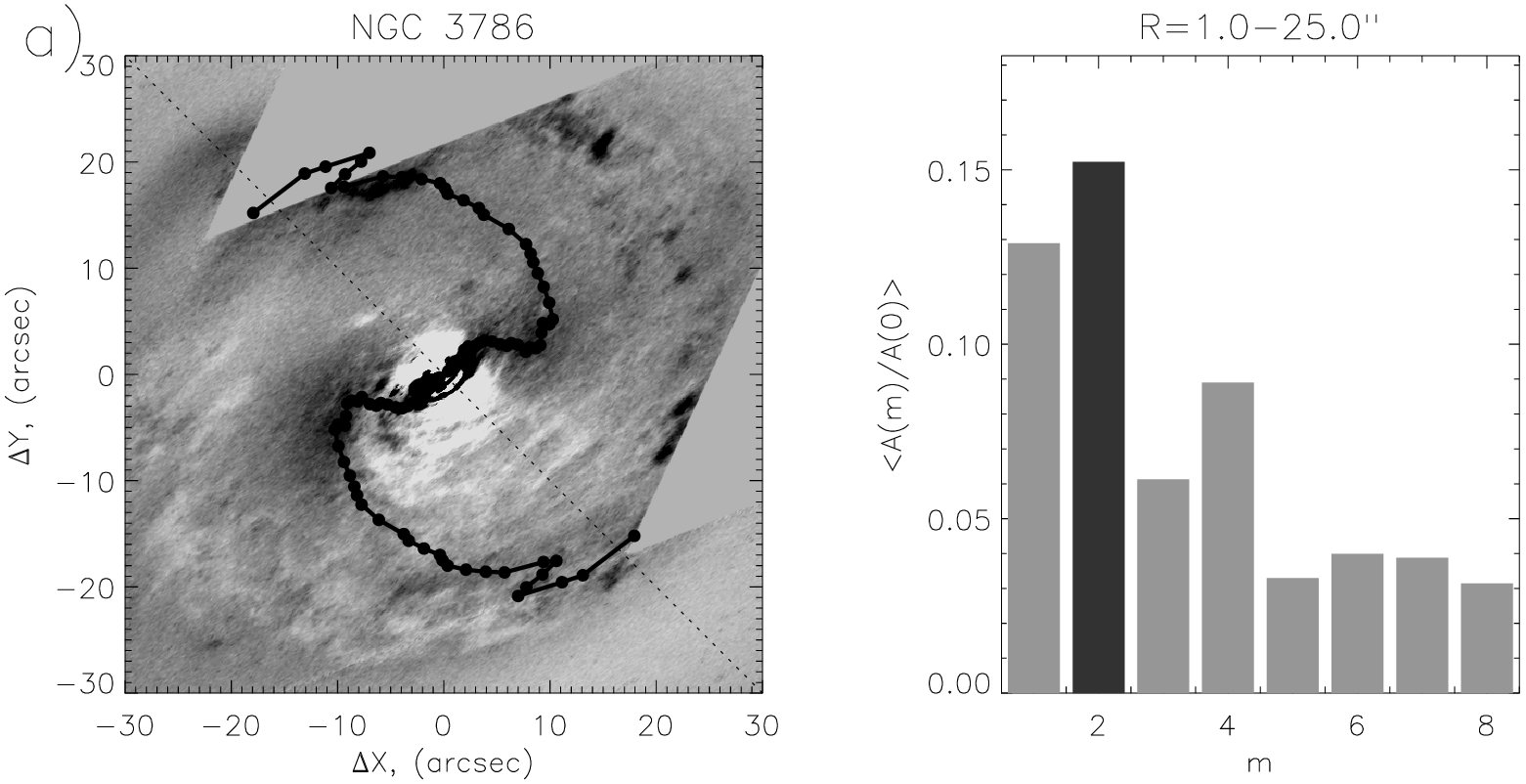}
\vspace{0.3 cm}

\includegraphics[width=15 cm]{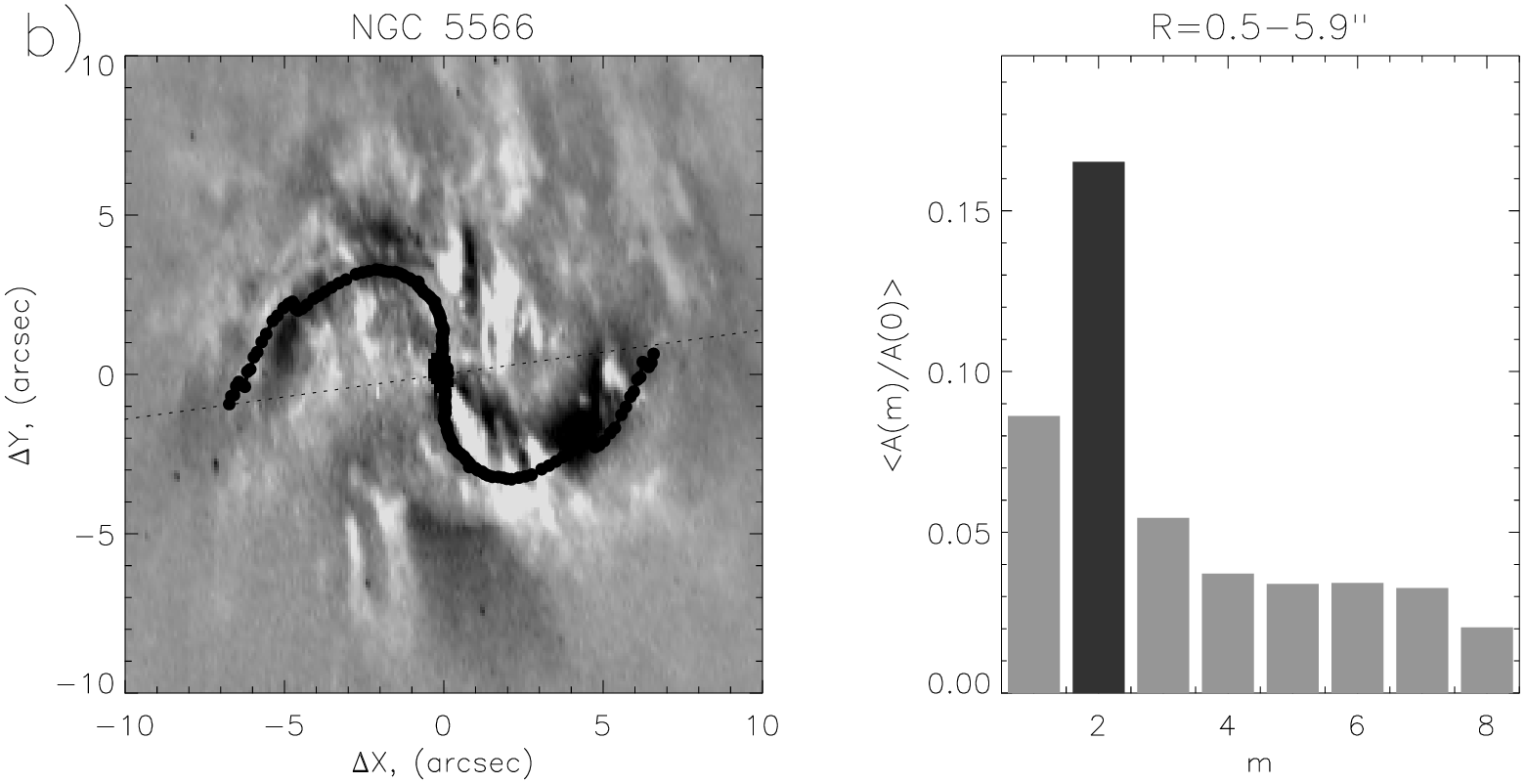}
\vspace{0.3 cm}

\includegraphics[width=15 cm]{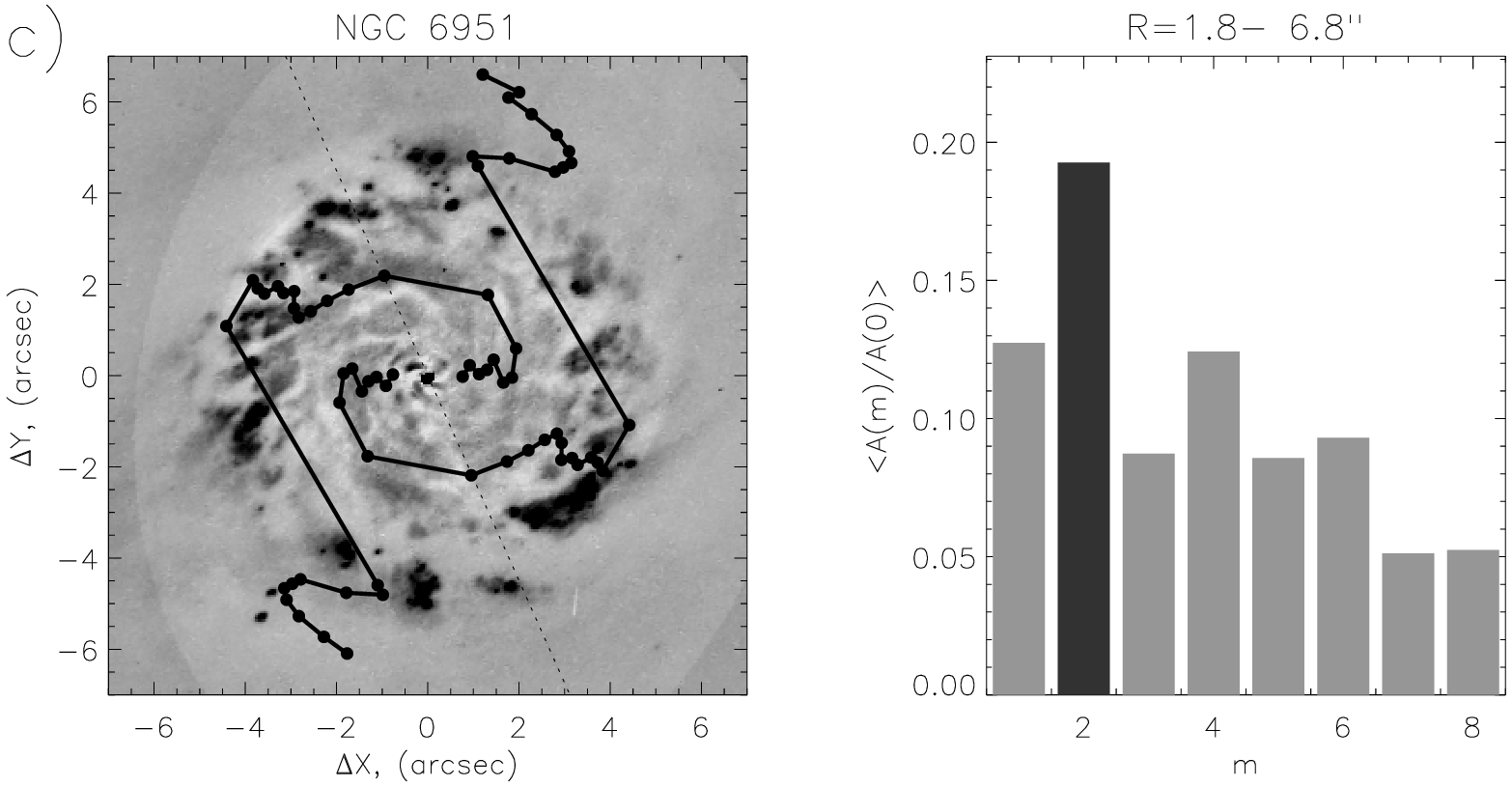}
 \protect\caption{
  The spiral patterns seen in the HST images (F606W filter):
(a) NGC3786, (b) NGC5566, and (c) NGC6951.  The maps of residual brightness in
the galactic plane with the lines of maximum of the second harmonic in the
brightness distribution superposed on them are shown on the left.  The mean
harmonic amplitude normalized to $A_0$ is shown on the right; the range of
radii ($R=..$) in which the averaging was performed is indicated. }
\end{figure*}

Three of the minispirals studied have already been described in the
literature: the pseudoring in NGC~2273 (Ferruit et al., 2000) and the
flocculent spirals in NGC~4736 (Elmegreen et al., 2002) and in NGC~6951
(P\'erez et al., 2000).  Two new circumnuclear spirals were also found: the
two-arm spiral in NGC~5566 and the three-arm spiral in NGC~7743.  Based on an
isophotal analysis of images, several authors (see references in the Table~1)
pointed out the existence of a secondary inner bar in these galaxies.  Within
the bar, the line of maximum of the principal $m = 2$ azimuthal harmonic must
be elongated along the corresponding constant azimuth.  In NGC~5566, however,
this line is wound into a spiral that runs almost from the very center and
coincides with the residual-brightness peak (Fig.~6b); i.e., a minispiral
rather than a bar is located here, with its shape resembling the outlines of
the spirals within the bars obtained in the model calculations by Englmaier
and Shlosman (2000).  The minispiral shown in Fig.~6b lies within a large scale
bar with a semi-major axis of about $30''$ ($\sim3$ kpc).

 In NGC~3786, the line of maximum of the $m = 2$
harmonic at $r < 5-6''$ has a constant azimuthal angle, which is a further
confirmation of the existence of a minibar whose dynamical manifestations are
described in Subsect.~3.2.  Far from the center, this line is wound into a
regular spiral that coincides with the global two-arm pattern in the galactic
disk at $r = 20-30''$.  The Fourier spectrum is dominated by the second
harmonic; the m = 1 harmonic is similar in amplitude to it (Fig.~6a). Clearly,
the large relative amplitude of the first harmonic can be explained in terms
of asymmetry in the apparent distribution of the dust lanes, most of which lie
on the near side of the NGC~3786 disk.  A similar effect was described by
Fridman and Khoruzhii (2000) when performing a Fourier analysis of the images
for NGC~157.  The spirals, defined as the maximum of the $m = 2$ harmonic near
the bar ends, change their winding direction (Fig.~6a). According to Fridman
and Khoruzhii (2000), this behavior must suggest that the bar is slow (in
terms of the angular velocity of rigid rotation).  For the nature of the
minibar in NGC~3786 to be eventually elucidated, we must know a more detailed
gas velocity field than that used here.

Wozniak et al. (1995) detected an isophotal twist at $r < 5-6''$ in the
ground-based optical images of NGC~6951 within its outer bar $r\approx60''$ in
size, which they interpreted as an inner bar.  However, based on near-infrared
photometry, Friedli et al. (1996) questioned this interpretation by assuming
that the isophotal twist could result from a complex distribution of dust and
star-forming regions within the central kiloparsec. Indeed, the HST images
exhibit a ring of star-forming regions that is elliptical in the plane of the
sky but is almost circular in the galactic plane (the inclination of the
galactic plane to the line of sight was assumed to be $42^\circ$, in agreement
with the data of P\'erez et al. (2000).  These high-resolution images confirm
the absence of a bar as an elongated, ellipsoidal structure.  The $m = 2$
harmonic dominates in the Fourier spectrum within the ring of star formation,
but, in contrast to NGC~3786 and NGC~5566, the line of its maximum consists of
separate fragments of the spirals.  This is also true for higher harmonics,
which, however, have a sufficiently high amplitude (Fig.~6c).  If we draw an
analogy with the  classification of large-scale spiral arms, then a grand
design is observed in NGC~3786 and NGC~5566 and flocculent spirals are observed
in NGC~6951 and NGC~4736.  However, the minispiral in NGC~6951 differs from the
large-scale spirals in disk galaxies in that the multi-arm spiral seen in the
residual-intensity distribution (Fig.~6c) is probably associated only with the
distribution of gas and dust rather than with the stellar component.  As was
shown by P\'erez et al. (2000), this inner spiral structure clearly seen in
the V band completely disappears in the H band, where the effect of dust
absorption is much weaker.

Interestingly, the position angle of the dynamical axis constructed from the
ionized-gas velocity field in the \Ha and [N~II] lines at $r = 0-8''$ deviates
by $10-15^\circ$ from the line of nodes of the outer disk, suggesting a
significant role of non-circular motions. Since there is no inner bar here, we
can offer the following interpretation of the observed pattern. There is a
gas-and-dust disk $6-8''$ ($400-600$ pc) in radius within the large-scale bar
in which a multiarmed spiral structure perturbing the circumnuclear gas
rotation develops. The dynamical decoupling of this disk is also confirmed by a
high molecular-gas density (Kohno et al. 1999) and by the location of two inner
Lindblad resonances of the large-scale bar here\footnote{When the paper was
submitted for publication, the paper by Rozas et al., (2002) appeared.  Based
on their  IFP observations, these authors reached a similar conclusion
regarding the nature of the inner disk in NGC~6951.} (P\'erez et al., 2000).

Presently, minispirals have been detected in the circumnuclear regions of many
galaxies (Carollo et al., 1998), but as yet no unequivocal interpretation of
the nature of their formation has been offered [see Elmegreen et al., (2002)
for more details].  One of the reasons why the theoretical interpretations have
failed seems to be the scarcity of reliable data on the observed kinematics of
such spirals. The papers on this subject are still few in number (Laine et
al., 2001; Schinnerer et al., 2002) and from this point of view, measurements
of the gas velocities at the center of NGC~6951 can be of interest in their
own right.

\section{Discussion}

The main motive for this study is the search for any common features in the
kinematics  of bar-within-bar structures in an attempt to prove that the
secondary bar is dynamically decoupled.  A similar attempt has recently been
also made by Emsellem et al. (2001), who presented the results of their study
of stellar motions in four southern-sky galaxies. Using the method of
``classical'' long-slit spectroscopy Emsellem et al., (2001) concluded that
the inner bars were decoupled in three objects based only on the existence of
a circumnuclear line-of-sight velocity peak in long-slit cuts. However, such
features can also be explained in terms of more natural factors, such as a
peculiar mass distribution in the disk and the bulge or non-circular motions
in the outer bar, without invoking the hypothesis of a secondary bar.
Observations of two-dimensional kinematics, which allow the pattern of
non-circular gas and stellar motions to be determined, could give a more
definitive answer.  The results obtained in this way appear all the more
unexpected.

First, the shape of the line-of-sight stellar velocity dispersion distribution
(Subsect.~3.1) is determined only by the outer bar and does not depend on the
relative position of the inner bar-like structure.  Second, either
non-circular motions typical of the outer bar or good agreement with circular
rotation are observed in the stellar velocity fields; the latter is, probably,
explained by the fact that here, the line-of-sight stellar motions in the bar
and the bulge are added together (Subsect.  3.2).  Finally, the ionized-gas
velocity fields everywhere point to the presence of noticeable non-circular
motions.  However, they either correspond to the outer bar (as suggested by
from analysis of the radial behavior of $PA_{dyn}$) or are associated with the
inner spiral structure (NGC~6951) or with another individual peculiar features
of the galaxy (NGC~470, NGC~3945, and NGC~5850). Thus, it turns out that the
secondary inner bar seen in the galaxy images does not affect the observed
kinematics of the gas and stellar components in all the sample objects.

This conclusion is in conflict with the popular opinion of a dynamically
independent secondary bar, which is based on analysis of isophotal shapes and
on model calculations (Sect.~1). Maybe the methods used here and the limited
spatial resolution do not allow the kinematic features of the small-scale
inner bar to be evaluated.  However, this is not the case,  because the
minibars in NGC~3368 and NGC~3786 do not differ in their apparent sizes from
the secondary bars in the remaining galaxies (see the Table~1), but the
features of non-circular gas and stellar motions associated with them are
clearly detected.  In this case, our photometric analysis suggests that there
is no outer bar in these two galaxies (Subsects. 3.2 and 3.3).

The three galaxies with the most peculiar features in the observed gas motions
should be considered separately.  Although the asymmetric $m = 1$ mode
developed within the bar of NGC~470 and the polar gaseous disk in NGC~5850 are
rare structures, they, nevertheless, were observed by several authors in other
objects as well (see Subsect.~3.2).  It was also noted in this subsection that
the two galaxies have close massive companions. Therefore, the assumption that
the features of their gas kinematics result from the interaction with the
companions appears reasonable enough.

According to Kuijken et al., (1996), the gaseous disks in lentucular (S0)
galaxies exhibit counter-rotation relative to the stars in $24\pm8\%$ of the
cases\footnote{In what follows, the errors are given at a $1\sigma$ level of
the binomial distribution.}; i. e. , this is a common phenomenon that is,
probably, attributable to merging of the fallen gaseous cloud with the
corresponding direction of angular momentum (Bertola et al., 1992).  Against
this background, the gas counter-rotation at the center of NGC~3945, one of
the four S0 galaxies in our sample (table~1) with gaseous disks observed in
three of them (NGC~2681, NGC 3945, and NGC~7743), comes as no surprise.

The presence of minispirals in some galaxies is not surprising either.  Thus,
Erwin and Sparke (2002) found that their sample of early-type (S0-Sa) barred
galaxies contained $24\pm7\%$ of objects with circumnuclear minispirals, which
is slightly less than $39\pm14\%$ (5 of 13) in the sample under study. Although
the difference between the frequencies of occurrence of minispirals is within
the error limits, it can be easily explained in terms of the selection effect
in the sample of galaxies with inner isophotal twists.  The inner spirals
distort the central isophotes, which may lead to the wrong conclusion that
there is a secondary bar.

Having studied the detailed kinematics of the gas and stars in the galaxies
from our sample, we may conclude that the so-called double-barred galaxies are
not a separate type of galaxies but are a combination of objects with
distinctly different structures of their circumnuclear regions.  The formal
use of isophotal analysis of images to study galaxies without invoking
kinematic data can lead to erroneous conclusions.  We think that the
double-barred galaxies described in the literature can be arbitrarily
divided   into two basic classes. The first class includes early-type (S0-Sa)
galaxies.  Here, the illusion of a secondary bar results from the triaxial
bulge shape.  In contrast to the bar, the triaxial bulge has virtually no
effect on the disk dynamics in the circumnuclear region.  A characteristic
example is NGC~2950. The second class includes galaxies of later types.  Here,
decoupled gas-and-dust disks with a minispiral structure distorting the
isophotal shape can be observed within large-scale bars.  A characteristic
example is NGC~6951.  There is also a third possibility.  The $x_2$ family of
stable orbits oriented perpendicular to the bar major axis can exist within a
large-scale bar (Contopouls and Grosbol, 1989).  A bar-like structure that is
exactly perpendicular to the primary bar and that correspondingly distorts the
observed isophotes can be formed on the basis of these orbits. Such a model
was proposed for NGC~2273 (Petitpas and Wilson, 2002).  In this case, however,
there is no decoupled secondary bar but there is a feature in the inner
structure of the major bar that rotates with it as a whole.

\section{Conclusions}

  A number of contradictions between existing models of nested bars and
observations can be resolved by abandoning the popular view of dynamically
independent double bars, which we propose here.  Many models suggest that the
secondary bar is a short-lived structure, which, in turn, is in conflict with
common observations of inner isophotal twists [see Erwin and Sparke (2002) for
more details].  However, all falls into place if these twists are not
associated with the secondary bars, at least in most of these galaxies.  On
the other hand, the fact that, according to statistical data, the photometric
secondary bar is in no way associated with the presence of an active nucleus
in the galaxy can be explained (Laine et al., 2002; Erwin and Sparke, 2002),
although it is clear from theoretical considerations that the correlation here
must be much closer than that for single bars.  The reason is that the
secondary bar traceable by isophotal twists is not a dynamically decoupled
structure in the galaxy at all.

\begin{acknowledgements}

I wish to thank V.L. Afanasiev for interest and helpful discussions and N.V.
Orlova for help in preparing the article.  This study is based on
observational data obtained with the 6m Special Astrophysical Observatory
telescope financed by the Ministry of Science of Russia (registration no.
01-43) and on HST NASA/ESA data taken from the archive of the Space Science
Telescope Institute operated by the Association of Universities for Research
in Astronomy under a contract with NASA (NAS 5-26555).  In my work, I used the
NASA/IPAC (NED) Extragalactic Database operated by the Jet Propulsion
Laboratory of the California Institute of Technology under a contact with the
National Aeronautics and Space Administration (USA) and the HYPERCAT database
(France).  The study was supported by the Russian Foundation for Basic
Research (project nos. 01-02-17597 and 02-02-06048mas) and the Federal Program
``Astronomy'' (project no.  1.2.3.1).
\end{acknowledgements}

\bigskip
{\it Translated by V. Astakhov}


\begin{thebibliography}{}

\bibitem[]{} Afanasiev V.L. \&  Shapovalova A.I., 1981, Astrophysics, 17, 221

\bibitem[]{} Afanasiev V.L. \&  Sil'chenko  O.K.,  1999, AJ,  117,  1725

\bibitem[]{} Afanasiev V.L.,  Mikhailov  V.P.,  Shapovalova  A.I.,   1998, Astron. Astrophys. Trans,  16, 257


\bibitem[]{} Afanasiev V.L., Dodonov S.N., Moiseev A.V., 2001, Stellar dynamics: from
classic to modern,  eds. Osipkov L.P., Nikiforov I.I., Saint Petersburg, 103

\bibitem[]{} Begeman  K.G.,  1989,  A\&A, 223,  47

\bibitem[]{} Bertola F., Buson L.M., Zeilinger W.W., 1992, ApJ,  401,  L79,

\bibitem[]{} Buta R.,  Crocker  D.A.,  1993, AJ,  105, 1344

\bibitem[]{}  Carollo C.M., Stiavelli M.,  Mack J., 1998, AJ, 116, 68

\bibitem[]{} Chevalier R.A., \& Furenlid I., 1978, AJ,  225,  67

\bibitem[]{}  Combes F., 2001, in Advanced Lectures on the Starburst- AGN Connection, Ed. by I.
Aretxaga, D. Kunth, \& R. Mejica (World Scientic, Singapore, 2001), 223,
(astro-ph/0010570)

\bibitem[]{}  Contopouls G., \&  Grosbol P., 1989,  A\&A Rev., 1, 261

\bibitem[]{}  de Vaucouleurs G., 1975, ApJs,  29,  193

\bibitem[]{}  Duval M.F., \&  Monnet G., 1985, A\&AS,  61, 141

\bibitem[]{}  Elmegreen D.M., Elmegreen B.G.,  Eberwein K.S, 2002, ApJ,    564, 234


\bibitem[]{}  Emsellem E., 2002, (astro-ph/0202522)

\bibitem[]{}  Emsellem E., Greusard D., Combes F., et al.,  A\&A, 2001, 368, 52

\bibitem[]{}  Englmaier P., \& Shlosman I., 2000, ApJ,  528, 677

\bibitem[]{}  Erwin P., \& Sparke L.S., 1999, ApJ,   521, L37

\bibitem[]{}  Erwin P., \& Sparke L.S., 2002, AJ, 124, 65

\bibitem[]{}  Ferruit P., Wilson A. S.,  Mulchaey J., 2000, ApJs,  128,  139

\bibitem[]{}  Frei Z., Guhathakurta P., Gunn J.E., Tyson J.A., 1996, AJ,  111, 174

\bibitem[]{}  Fridman A.M., \&  Khoruzhii O.V., 2000,  Phys. Lett. A, 276, 199

\bibitem[]{}  Fridman A. M.,  Khoruzhii O. V.,  Polyachenko E. V., et al., 2001b,  MNRAS,  323,  651

\bibitem[]{}  Friedli D., \&  Martinet L., 1993, A\&A,  277, 27

\bibitem[]{}  Friedli D., Wozniak H., Rieke M., Martinet L., Bratschi P., 1996, A\&AS,  118, 461

\bibitem[]{}  Heller C., Shlosman I., Englmaier P.,  2001, ApJ,   553, 661

\bibitem[]{}  Higdon J., Buta R., Purcel G.B., 1998, AJ,  115,  80

\bibitem[]{}  Jungwiert B., Combes F., Axon D.J., 1997, A\&AS,  125, 479

\bibitem[]{}  Khoperskov A.V.,  Zasov A.V.,  Tyurina N.V., 2002, Astron. Reports. 45, 180

\bibitem[]{}  Khoperskov A.V., Moiseev A.V., Chulanova E.A., 2001, Bull. Spec. Astrophys. Obs.,  52, 135

\bibitem[]{}  Knapen J.H., Shlosman I., Heller C.H., et al., 2000a, ApJ, 528, 219

\bibitem[]{}  Knapen J.H., Shlosman I.,  Peletier R.F, 2002b, ApJ. 529, 93

\bibitem[]{}  Kohno K., Kawabe R., Vila-Vilaro B., 1999, ApJ,   511,  157

\bibitem[]{}  Kormendy J,, 1982, ApJ, 257, 75

\bibitem[]{}  Kormendy J,, 1983, ApJ, 275, 529

\bibitem[]{}  Kuijken K., Fisher D., Merrifield M.R., 1996, MNRAS,  283,  543

\bibitem[]{}  Laine S.,  Knapen J.H., P\'erez-Ramyrez D., et al., 2001, MNRAS, 324, 891

\bibitem[]{}  Laine, S., Shlosman, I., Knapen, J.H., Peletier R.F., 2002, ApJ, 567, 97

\bibitem[]{}  Lindblad P.O., 1999,  A\&A Rev. 9, 221

\bibitem[]{} Lyakhovich V.V., Fridman A.M., Khoruzhii O.V., Pavlov A.I., 1997, Astron. Reports, 41, 447

\bibitem[]{}  Maciejewski W., Sparke L.S.,  2000,  MNRAS,   313 745

\bibitem[]{}  W. Maciejwski, J. Teuben, L. S. Sparke, \& J. M. Stone, Mon. Not. R.
Astron. Soc. 329, 502 (2002).

\bibitem[]{}  Maciejewski W., Teuben J., Sparke L.S., Stone J.M., 2002,  MNRAS,  329  502

\bibitem[]{}  Moiseev A.V., 2001a, Bull. Spec. Astrophys. Obs.,  51, 11 (astro-ph/0111219)

\bibitem[]{}  Moiseev A.V., 2001b, Bull. Spec. Astrophys. Obs.,  51, 11 (astro-ph/0111219)

\bibitem[]{}  Moiseev A.V., 2002, Bull. Spec. Astrophys. Obs., 54, (astro-ph/0211104)

\bibitem[]{}  Moiseev,  A.V.,  \& Mustsevoi  V.V,  2000,  Astro.  Letter., 26,  665 (astro-ph/0011225)

\bibitem[]{}  Moiseev, A.V., Valdes, J.R., Chavushan, V.O., 2002,  Preprint SAO RAS, No. 171,
submitted to A\&A

\bibitem[]{}  Monnet G., Bacon R., Emsellem E., 1992, A\&A,   253,  366

\bibitem[]{}  Mulchaey J.S., Regan M.W., Kundu A., 1997,  ApJs,  110,  299

\bibitem[]{}  P\'erez E., M\'arquez I., Durret F., et al., 2000, A\&A,  353, 893

\bibitem[]{}   Petitpas  G.R., \& Wilson C.D., 2002, ApJ, 575, 814

\bibitem[]{}  Pfenninger, D., \& Norman, C.A.,  1990, ApJ,  363, 391

\bibitem[]{}  Rozas M.,  Relano M.,  Zurita A.,  Beckman J. E., 2002, A\&A,  386 42

\bibitem[]{}  Sakamoto K.,  Okumura S.K., Ishizuki S.,  Scoville N.Z., 1999, ApJ, 525, 691

\bibitem[]{} Schinnerer E., W. Maciejewski W.,  Scoville N.,  Moustakas L.A., 2002, ApJ, 575, 826

\bibitem[]{}  Selwood J.A., \& Wilkinson A., 1993, Rep. Prog. Phys. 56, 173

\bibitem[]{}  Shaw M.A., Combes F., Axon D.J., Wright G.S., 1993, A\&A,  273, 31

\bibitem[]{}  Shlosman I., Frank J., Begeman M.C., 1989, Nature,  338, 45

\bibitem[]{}  Shlosman I., \& Heller C., 2002, ApJ,  565,  921


\bibitem[]{}  Sil'chenko O.K., Burenkov A.N., \&  Vlasyuk V.V., 1997, A\&A,  326, 941

\bibitem[]{}  Sil'chenko O.K., 2002, Astron. Lett., 28, 207


\bibitem[]{} Sil'chenko, O.K., Moiseev, A.V., Afanasiev, V.L., Vald\'es J.R., Chavushan
V.O., 2003,  submitted to AJ

\bibitem[]{}  Turnbull, A.J.,  Bridges, A.J., Carter, D., 1997, MNRAS,   307,  967

\bibitem[]{}  Vauterin P., \& Dejonghe H., 1997, MNRAS,  286, 812

\bibitem[]{}  Wozniak H.,  Friedli D., Martinet L., Martin P., Bratschi P., 1995, A\&AS,  111, 115

\bibitem[]{}  Zasov, A.V. \& Khoperskov, A.V., 2002, Astronomy Reports,  46, 173


\end{thebibliography}
\end{document}